\documentclass[aps,prx,twocolumn,showpacs,amsmath,amssymb,longbibliography, amsfonts,10pt,superscriptaddress]{revtex4-2}
\usepackage{graphicx} 
\usepackage{amsmath}
\usepackage{amsfonts}
\usepackage{amssymb}
\usepackage{graphicx}
\usepackage{color}
\usepackage{xcolor}
\usepackage{bbold}
\usepackage{appendix}
\usepackage[citecolor=blue]{hyperref}
\hypersetup{colorlinks}
\usepackage{subcaption}
\usepackage{bookmark}
\usepackage{makecell}
\usepackage{mathtools}
\usepackage{comment}
\usepackage{tikz}
\usetikzlibrary{arrows.meta}

\usepackage{array}
\usepackage{multirow}
\usepackage[nolist,nohyperlinks]{acronym}
\usepackage[export]{adjustbox}

\newcommand{\vecr}{\boldsymbol{r}}
\newcommand{\vecq}{\boldsymbol{q}}

\newcommand{\vecxi}{\boldsymbol{\xi}}
\newcommand{\vecphi}{\boldsymbol{\phi}}
\newcommand{\vecj}{\boldsymbol{j}}
\newcommand{\vecv}{\boldsymbol{v}}
\newcommand{\vecA}{\boldsymbol{A}}

\begin{document}
\title{Meissner Effect and Josephson Radiation in Driven Dissipative Superconductors}
\author{Carl Philipp Zelle}
\affiliation{Department of Physics, Harvard University, Cambridge MA 02138, USA}
\author{Subir Sachdev}
\affiliation{Department of Physics, Harvard University, Cambridge MA 02138, USA}
\affiliation{Center for Computational Quantum Physics, Flatiron Institute, 162 5th Avenue, New York, NY 10010, USA}
\begin{abstract}
    Photo-induced superconducting signatures have been observed in several materials at temperatures far above their equilibrium critical temperatures, including magnetic field expulsion as in the Meissner effect. We propose driven-dissipative condensation of the superconducting order parameter as a mechanism for this phenomenon. In such a theory, optical pumping generates an effective gain (possibly via a parametric resonance) that overcomes the intrinsic damping of the pairing field, and stabilizes a non-equilibrium condensate whose phase rotates at an intrinsic frequency that is generally lower than, and incommensurate with, the drive frequency. We develop a phenomenological continuum theory that couples this slowly rotating order parameter to the conserved charge density and the electromagnetic gauge field. Despite its finite-frequency dynamics, the resulting state exhibits the conventional long-wavelength electromagnetic signatures of superconductivity. In three dimensions, it displays a static Meissner effect and a gapped plasmon generated by the Anderson--Higgs mechanism. In a two-dimensional sheet, it exhibits Pearl screening and the characteristic square-root plasmon dispersion. We further propose a direct experimental test based on a Josephson junction between the driven-dissipative state and an equilibrium superconductor. The junction supports an AC Josephson current at zero applied voltage and emits radiation at the intrinsic rotation frequency of the condensate. Its low, pump-dependent, and generally incommensurate frequency provides a clear signature distinguishing driven-dissipative superconductivity from equilibrium and drive-locked pairing states.
\end{abstract}
\maketitle

\section{Introduction}

Using light to create and control phases of quantum matter has emerged as a powerful route to material properties that are inaccessible in equilibrium.
A striking recent example has been the observation, by Fava {\it et al.\/} \cite{fava2024}, of magnetic flux expulsion similar to the Meissner effect in an optically driven cuprate superconductor at temperatures ($T$) well above the equilibrium $T_c$; several other experiments have detected other indications of superconductivity 
\cite{fausti2011,mitrano2016,cremin2019,Shimano23,Shimano24,Xu25}. Also notable are observations of similar signatures in a quantum spin liquid \cite{mitrano2026}.

\begin{figure*}
    \centering
    \subfloat[\label{fig:incoherent_pumpA}]{
    \includegraphics[width=0.3\linewidth]{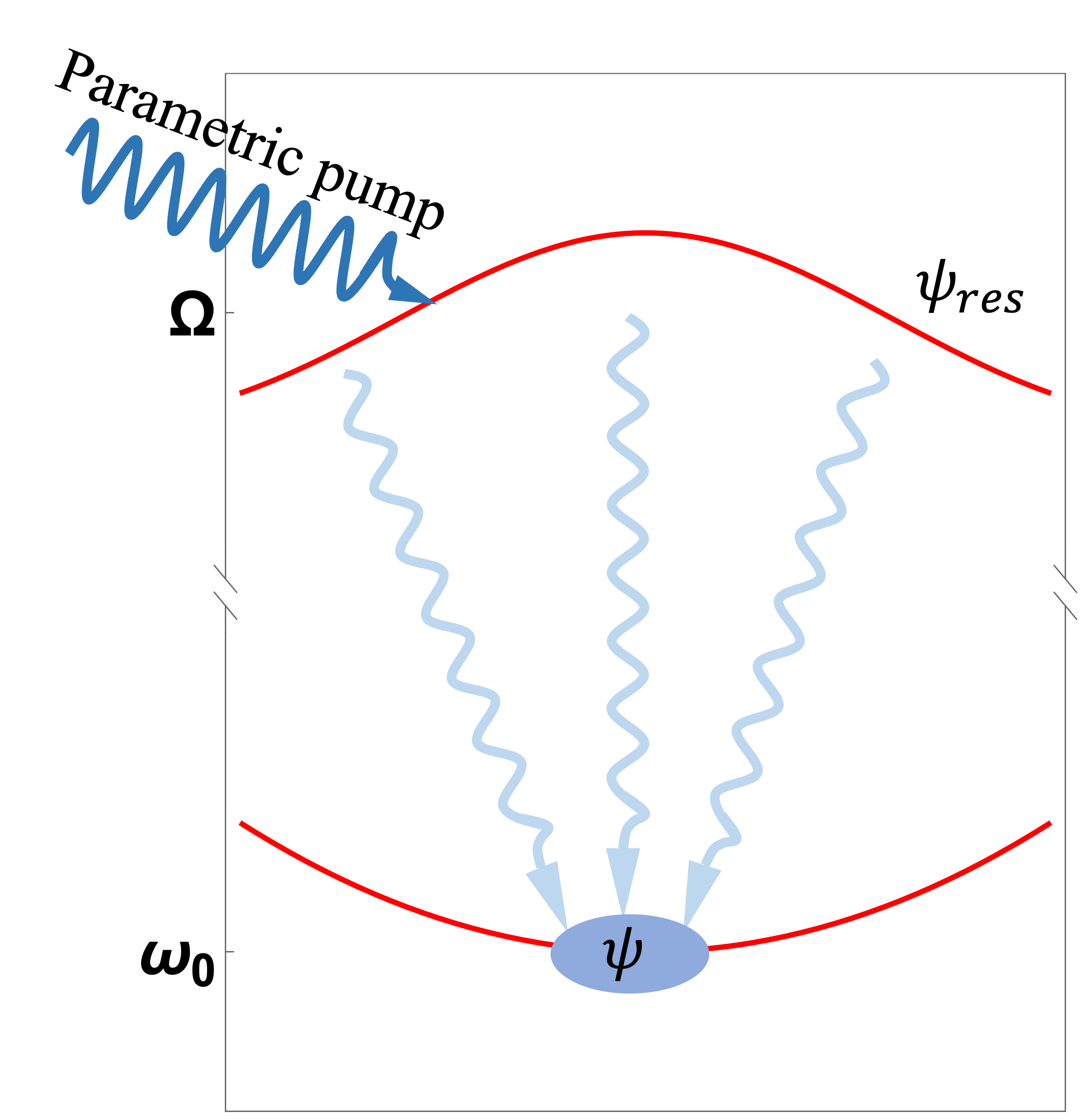}}
    \subfloat[\label{fig:incoherent_pumpB}]
    {\includegraphics[width=0.4\linewidth]{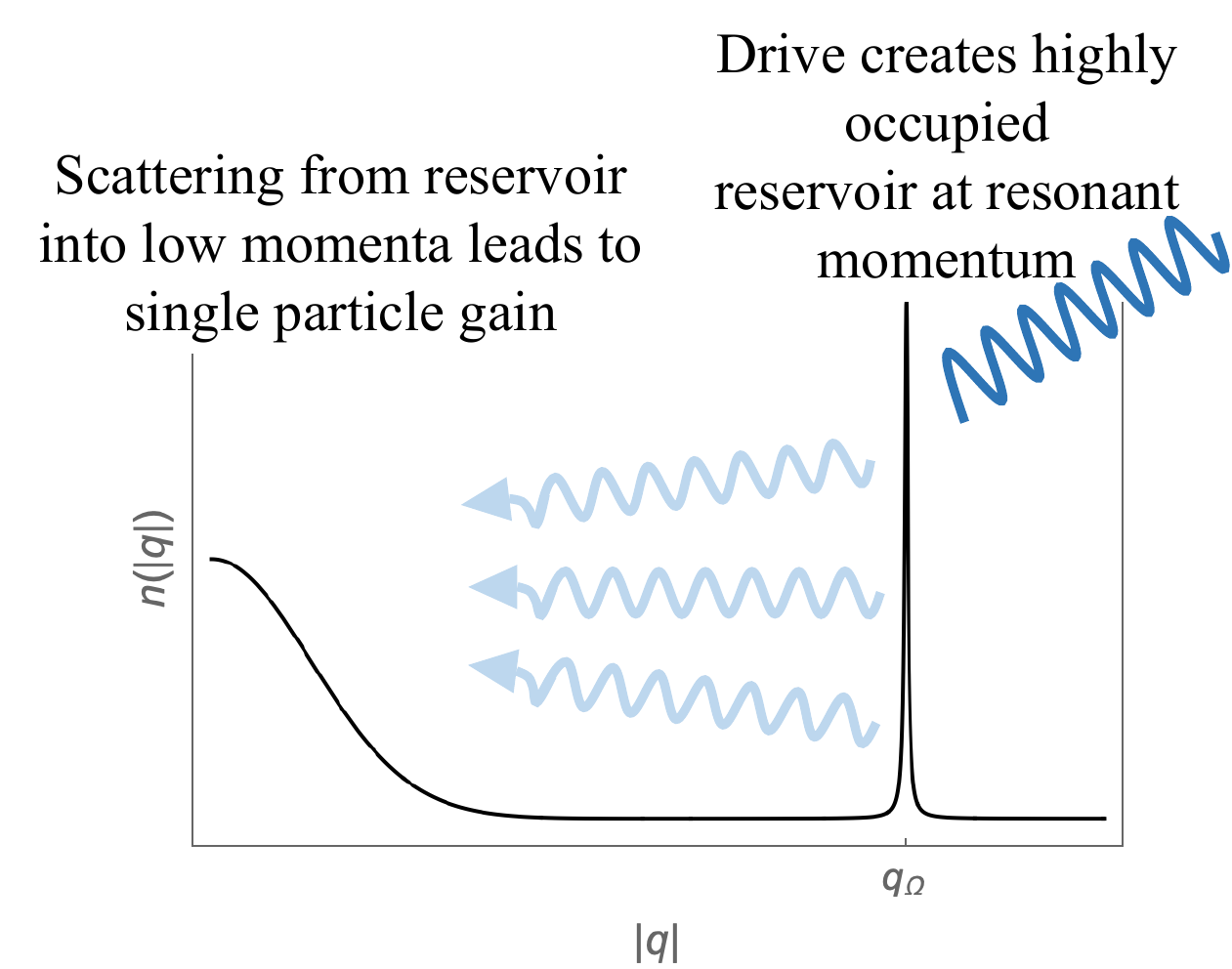}}
    \subfloat[\label{fig:phaseDiag}]
    {\includegraphics[width=0.3\linewidth]{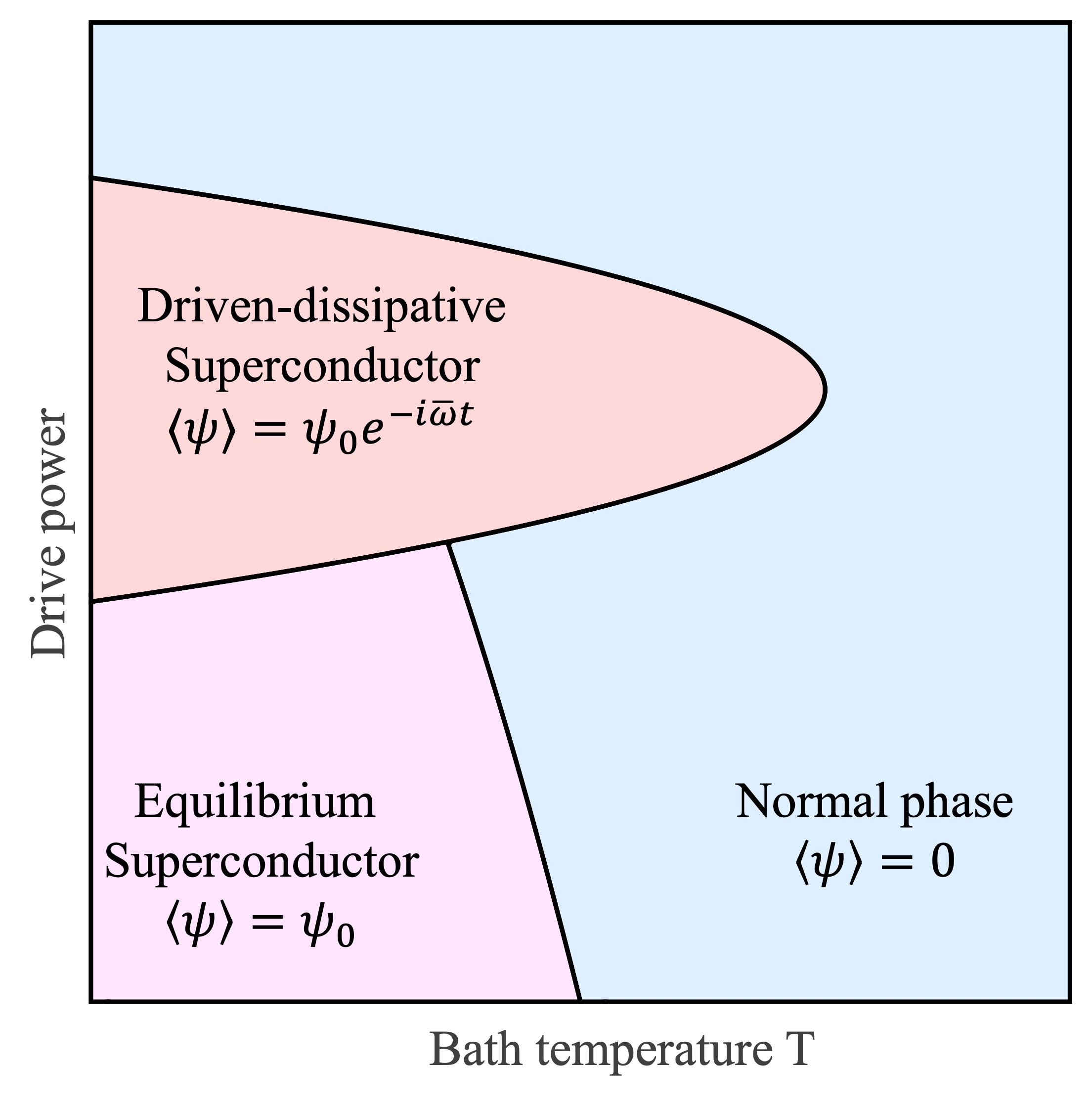}}
    \caption{Left: schematic incoherent pumping scheme. Laser drive at a high frequency $\Omega$ creates occupations at a reservoir degree of freedom $\psi_{res}$, possibly via parametric resonance. These excitations quickly relax through collisions into the low frequency regime. This leads to an effective negative contribution to the dissipation of the slow degrees of freedom. For sufficiently strong pumping, a condensate rotating at a low frequency $\omega_0$ forms. Middle: For parametrically driven field theories high momentum modes can serve as the pumped reservoir. These occupations relax through collisions into the long wavelength regime where they can induce a dissipative instability. Right: Phase diagram of a driven dissipative superconductor. There is the usual ordering transition upon cooling the bath down. Pumping can cause a dissipative condensation of a slowly rotating pairing field. Adapted from Ref.~\cite{zelle2026}.}
    \label{fig:driven_order}
\end{figure*}

These observations have spurred a series of theoretical proposals to explain this phenomenon \cite{kennes2017,eckhardt2024,chattopadhyay2025,Michael2025,diessel2025,chattopadhyay2026,diessel2026}  building upon earlier results \cite{Scalapino78,Knap15,Komnik16,Chandran16,Babadi17,DaiLee21,Mitra21a,Mitra21b,Dai22,Sols24,Chandra25a,Moessner25,Chandra25b}. 
These works rely on a parametric resonance (sometimes called a `discrete time crystal') to produce an amplification of the superconducting order parameter oscillating at a frequency $\Omega/2$, where $\Omega$ is the frequency of the pump
(see Fig.~\ref{fig:phaseDiag}).

Here, we use recent arguments \cite{zelle2024,daviet2024,zelle2026,okugawa2026} that such driven oscillating many-body states are generically unstable to collisions between the excitations. Nevertheless, the long-time fate is not generically a fully disordered state. Under suitable conditions, there can be a phase transition to a state in which there is a globally-averaged non-zero order parameter which is coherently slowly rotating at frequency significantly smaller than the drive. The underlying physics of this rotating state is {\it not\/} a parametric resonance, but a change in sign of the effective damping introduced by the optical pumping. 

So we are proposing that photo-induced superconductivity arises from a dissipative instability of the superconducting order parameter.
It has recently been shown, that optical drives can cause exactly this type of instability \cite{zelle2026, okugawa2026}.  As shown in Fig.~\ref{fig:driven_order}, occupation of high energy modes, possibly via a parametric amplification \cite{vonHoegen2022,Michael2020}, is followed by collisions which lead to the formation of a condensate at low energy.
This anti-damping instability occurs at low frequencies that are set by the material itself, and are in general much lower and incommensurate with the drive. The resulting phase does show long range order which is however dynamical: It slowly rotates and therefore constitutes a far-from equilibrium stable state that cannot be part of any equilibrium phase diagram \cite{zelle2024,daviet2024}. A schematic non-equilibrium phase diagram in shown in Fig.~\ref{fig:phaseDiag}.

The mechanism behind the antidamping instability is reminiscent of incoherent pumping in polariton platforms where it leads to the phenomenon of driven dissipative superfluidity or Bose-condensation \cite{imamoglu1996,kasprzak2006,balili2007,deng2002,Fontaine2021,byrnes2014,deng2010,Carusotto2013}. The basic mechanism is the following: a high frequency laser pump coherently generates excitations of a degree of freedom that is at resonance with the drive. These excitations immediately start to relax into the lower frequency degrees of freedom through collisions. From the low frequency regime's point of view this manifests as a steady incoherent gain rate of excitations. The resonantly pumped high frequency modes constitute a reservoir with inverted occupation. If this incoherent gain exceeds the inherent loss or dissipation of the low frequency mode this triggers the driven-dissipative condensation.

Here, we develop a phenomenological continuum theory of driven-dissipative superconductivity and determine its electromagnetic response. 
In Section~\ref{sec:driven}, we review the general mechanism of driven-dissipative condensation of a complex order parameter, and then extend it to include a conserved charge density and the electromagnetic gauge field. In Section~\ref{sec:Meissner}, we show that the resulting state exhibits a static Meissner response, characterized by a London penetration depth in three dimensions, and a Pearl screening length in a two-dimensional sheet. The Anderson–Higgs mechanism also survives: bulk plasmons are gapped, whereas plasmons in a two-dimensional sheet retain their characteristic square-root dispersion. In Section~\ref{sec:josephson}, we propose a Josephson-junction experiment that distinguishes this state from both an equilibrium superconductor and a resonantly driven pairing field. The intrinsic rotation of the order parameter at constant amplitude produces an AC Josephson current at zero applied voltage, with a frequency that is generally lower than and incommensurate with the optical drive. We note that for the case of a pairing field whose amplitude oscillates at the high time-crystal frequency $\Omega/2$ a similar Josephson response was discussed in Ref.~\cite{Dai22}.

\section{Driven-Dissipative condensation}
\label{sec:driven}

To start we consider a matter system with a bosonic gap $\omega_0$. It is coupled to a cool bath which generally gives rise to dissipation and therefore finite life-times of the excitations in the system. We assume that the life-time of the lowest lying excitation is the largest in the system. We call these long lived excitations $\psi(t,\vecr)$. Furthermore we assume that the system also has resonances at much larger gap values $\sim \Omega\gg\omega_0$ which we call $\psi_{res}(t,\vecr)$. The exact excitation spectrum of the system does not matter at this point. We now drive this system with a laser at frequency $\Omega$ such that the high frequency matter degrees of freedom $\psi_{res}$ are resonantly pumped. The thereby created excitations immediately relax and through collision processes scatter into the low frequency regime. These collision processes are generally irreversible, meaning that the reservoir modes $\psi_{res}(t,\vecr)$ feed into the the long lived regime $\psi(t,\vecr)$ but not vice versa. This leads to a continuous, incoherent effective gain rate at which excitations drop into the long-lived matter degrees $\psi(t,\vecr)$ while the coupling to the cold thermal bath prevents the system from heating up to infinite temperature. If the gain rate is large enough, this can lead to an instability in the \emph{dissipation} of $\psi(t,\vecr)$ causing so-called \emph{driven-dissipative condensation}.

The driven-dissipative condensate is an example of a true non-equilibrium steady state with long range order. In distinction from equilibrium order it rotates at a finite angular velocity that is set by the gap $\langle\psi(t,\vecr)\rangle=\psi_0e^{-i\bar\omega t},\,\bar\omega=\omega_0+\text{interaction corrections}$. The most well-known example of driven dissipative condensation is exciton-polariton condensation in semiconductor quantum wells immersed in optical cavities \cite{imamoglu1996,kasprzak2006,balili2007,deng2002}. In such systems there is two polariton branches, the upper one serves as the reservoir $\psi_{res}$ that is pumped by an external laser and the condensation ultimately occurs in the lower branch, see Fig.~\ref{fig:incoherent_pumpA} giving rise to a driven-dissipative superfluid \cite{Wouters2007,Carusotto2013,Keeling2011}. The mechanism of incoherent pumping however applies more generally. Recently, it was shown that rapid parametric drives of an $O(N)$ order parameter density can also induce a dissipative instability \cite{zelle2026}, see Fig.~\ref{fig:incoherent_pumpB}. In the case of an $O(2)$ order parameter $\vec \phi\in\mathbb{R}^2$ the driven open system therefore has three phases. There is two phases that also exist in the undriven system, a normal phase $\langle\vec\phi\rangle=0$ and an ordered (or superconducting) phase with $\langle\vec \phi(t)\rangle$. The driven system also displays a phase unique to the non-equilibrium whose order is given by a driven-dissipative condensate $\vec \phi = \phi_0(\cos{\bar\omega t},\sin{\bar\omega t})^T$. The resulting phase diagram was shown in Fig.~\ref{fig:phaseDiag}.

These results - the experimental observation of driven-dissipative or photo-induced superfluidity and the theoretical prediction of driven-dissipative order in driven $O(N)$ models - suggest that a superconducting order parameter can also undergo such a transition, when driven by a fast laser. This is for instance the case in the experimental observation of photo-induced superconducting behavior in various unconventional superconductors way above their respective critical temperatures in for example K$_3$C$_{60}$ \cite{mitrano2016}, the cuprates \cite{fausti2011,cremin2019,Shimano23,Shimano24,fava2024,Michael2025}, and the nickelates \cite{Xu25}. Here we develop a phenomenological continuum model of driven-dissipative superconductivity. A microscopic theory for unconventional compounds such as the cuprates is beyond the scope of this article. Such phenomenological field theories have successfully predicted e.g the excitation spectrum, transport behavior as well as Kardar-Parisi-Zhang (KPZ) scaling in the driven dissipative superfluids \cite{Wouters2007,Keeling2011, Altman2015, Zamora2017} that were observed experimentally in polaritons \cite{Roumpos2012,Claude2025,Fontaine2021, widmann2026}.

There are two important differences with respect to the driven-dissipative superfluid that a model of driven-dissipative superconductivity has to take into account. First, in driven-dissipative superfluids particle number is not conserved. And while the superconducting order parameter alone is also not a conserved quantity, the total charge $\nu$ is. This leads to an additional hydrodynamic mode that couples to the dynamics. This leads to a non-equilibrium version of Model F of Halperin and Hohenberg \cite{Hohenberg1977}
\begin{align}\label{eq:EoM_kartesian}
  &  \left(\partial_t+r-Z\nabla^2+\lambda|\psi|^2+g \nu\right)\psi+\xi=0\\
    &\partial_t\nu=\nabla \cdot(\vecj_\psi+\vecj_n)
\end{align}
where $\psi(t,\vecr)$ is the superconducting order parameter, and $r=i\omega_0+\gamma$ captures its gap $\omega_0$ as well as the effective dissipation $\gamma$. 
The dissipative instability towards non-equilibrium superconductivity occurs when $\gamma$ is tuned through zero. $Z=iZ_c+Z_d$ captures the momentum dependence of gap and dissipation, $\lambda=i\lambda_c+\lambda_d$ nonlinearities and $\xi$ is a complex Gaussian white noise with width $D$, 
\begin{align}
\langle \xi^*(t,\vecr)\xi(t',\vecr')\rangle=2D\delta(t-t')\delta(\vecr-\vecr')\,,     
\end{align}
capturing unavoidable finite temperature effects. Possible nonmarkovian colored contributions due to the time dependence of the drive are subleading in the long wavelength regime \cite{Bingyu2018,Gamba2025}.   The coupling of $\psi$ to the conserved charge density $\nu$ is captured by  $g=ig_c+g_d$. Its own dynamics follows a continuity equation with the supercurrent $\vecj_\psi$ and the current of the remaining charge carriers $\vecj_n$. In a phase amplitude decomposition $\psi(t,\vecr)=\sqrt{\rho(t,\vecr)}\exp{i\theta(t,\vecr)}$ the superconducting current reads $\vecj_{\psi}=Z_c\rho\nabla\theta$. We write the current of the remain non-condensed charge carriers $\vecj_n$ phenomenologically as $\vecj_n=n_f\vecv$. We assume a uniform density of carriers $n_f$ and the equation of motion for the velocity field $\vecv$ is
\begin{align}\label{eq:eom_charge}
    m\partial_t\vecv+\gamma_n\vecv+\alpha\nabla\theta+\vecxi_v=0.    
\end{align}
Here, $m$ is the mass of the individual charge carriers, $\gamma_n$ is the friction of the normal charge carriers, and $\alpha\neq0$ models a drag from a superfluid current that the normal charge carriers feel; $\vecxi_v$ is the corresponding thermal white noise.

The non-equilibrium nature of Eq.~(\ref{eq:EoM_kartesian}) resides in  the complex values of the couplings $Z$, $\lambda$, and $g$. Nevertheless, 
an effective thermal equilibrium in a co-rotating frame can be defined if \cite{Sieberer2015}
\begin{align}\label{eq:thermal_equilibrium}
    \frac{Z_c}{Z_d}=\frac{\lambda_c}{\lambda_d}=\frac{g_c}{g_d}.
\end{align}
In the non-equilibrium superconductor
\begin{align}\label{eq:saddle}
    \psi_0=\sqrt{\rho_0}e^{-i\bar\omega t},\quad\rho_0=\frac{-\gamma}{\lambda_d},\quad \bar\omega=(r_c+\lambda_c\rho_0).
\end{align}
We assume a neutralizing background such that the total charge density $\int_{\vecr}\nu(t,\vecr)=0.$ 

The second important difference is that $\psi$ is charged and therefore couples to the electromagnetic gauge field. This means that all derivatives acting on $\psi$ need to be replaced with covariant derivatives $\partial_\mu\rightarrow D_\mu=\partial_\mu-iA_\mu$ and adding the electric force $F=\nabla A_0+\partial_t \vecA$ to Eq. \eqref{eq:eom_charge}. Maxwell equation in Coulomb gauge $\nabla\cdot\vecA=0$ then read
\begin{align}
    \nabla^2A_0=n,\quad (\partial_t^2-c^2\nabla^2)\vecA=-2Z_c\rho \vecA+n_f\vecv_T,
\end{align}
with $\vecv_T$ the transversal component of the carrier velocity $\vecv$.

\section{Plasmon and Meissner effect}
\label{sec:Meissner}

We now determine the collective electromagnetic response of the driven-dissipative superconducting state, considering both a three-dimensional bulk superconductor and a two-dimensional superconducting sheet. The calculation proceeds by separating amplitude and phase fluctuations of the complex pairing field. Because the amplitude mode is gapped, it can be eliminated at frequencies and momenta below its relaxation scale. This yields an effective equation for the phase, which we combine with charge conservation, the dynamics of the normal carriers, and Maxwell's equations. The longitudinal sector determines the plasmon spectrum, whereas the transverse sector determines the magnetic response and the associated screening length.

It is useful to write the complex pairing field in phase-amplitude form,
\begin{align}
\psi=\sqrt{\rho}e^{i\theta}.
\end{align}
Equation~\eqref{eq:EoM_kartesian} then becomes
\begin{widetext}
\begin{align}\label{eq:eom_amplitude}
\partial_t\rho+2\gamma\rho+2\lambda_d\rho^2+2g_d\nu\rho-Z_d\left[\nabla^2\rho2\rho(\nabla\theta-\vecA)^2-\frac{(\nabla\rho)^2}{2\rho}\right]+2Z_c\nabla\cdot\left[\rho(\nabla\theta-\vecA)\right]+\xi_\rho&=0,
\\
\partial_t\theta-A_0-\omega_0-\lambda_c\rho-g_c\nu+\frac{Z_c}{2\rho}\left[\nabla^2\rho-2\rho(\nabla\theta-\vecA)^2-\frac{(\nabla\rho)^2}{2\rho}\right]-\frac{Z_d}{\rho}\nabla\cdot\left[\rho(\nabla\theta-\vecA)\right]+\xi_\theta&=0,
\end{align}
\end{widetext}
where $\xi_\rho$ and $\xi_\theta$ are the corresponding real-valued projections of the noise which are to leading order again Gaussian and white.
We expand around the spatially homogeneous rotating saddle point introduced in Eq.~\eqref{eq:saddle},
\begin{align}
\rho=\rho_0+h,\qquad\theta=\Omega t+\phi.
\end{align}
Here, $h$ denotes amplitude fluctuations, $\nu$ charge-density fluctuations, and $\phi$ phase fluctuations around the uniformly rotating state.

The amplitude fluctuation $h$ has a finite relaxation rate $2\lambda_d\rho_0$ at zero momentum and is therefore nonhydrodynamic. Expanding to linear order in $h$ and $\nu$, while retaining the leading nonlinear phase-gradient term, gives
\begin{equation}
\begin{split}
&\left(\partial_t+2\lambda_d\rho_0-Z_d\nabla^2\right)h+2Z_d\rho_0(\nabla\phi-\vecA)^2\\
&-2Z_c\rho_0\nabla^2\phi+2g_d\rho_0\nu+\xi_\rho=0.
\end{split}
\end{equation}
At frequencies and momenta small compared with the amplitude-relaxation scale, the time and gradient dependence of the inverse amplitude propagator may be neglected to leading order. Adiabatically eliminating $h$ then yields
\begin{align}\label{eq:eliminate_amplitude}
h=-\frac{Z_d(\nabla\phi-\vecA)^2-Z_c\nabla^2\phi+g_d\nu}{\lambda_d}+\frac{\xi_\rho}{2\lambda_d\rho}.
\end{align}

A leading in long-wavelengths expansion of the phase equation gives,
\begin{align}\label{eq:phase}
\partial_t\phi-A_0-\lambda_ch-g_c\nu-Z_c(\nabla\phi-\vecA)^2-Z_d\nabla^2\phi+\xi_\theta=0.
\end{align}
The resulting phase dynamics must be supplemented by the dynamics of the conserved charge density. The longitudinal normal current contributes to charge redistribution and therefore enters the continuity equation together with the condensate current.

The longitudinal component of the normal-carrier velocity is obtained from Eq.~\eqref{eq:eom_charge} as
\begin{align}
\vecv_L=\hat\sigma\left(\alpha\nabla\theta-\nabla A_0+\vecxi_L\right),\quad\hat\sigma=(m\partial_t-\gamma)^{-1}.
\end{align}
Assuming a spatially uniform density $n_f$ of normal carriers and inserting this expression into the continuity equation gives
\begin{align}\label{eq:charge}
\partial_t\nu-\left(Z_c\rho_0+\alpha n_f\hat\sigma\right)\nabla^2\phi-n_f\hat\sigma\nabla^2A_0=0.
\end{align}

Up to this point, the derivation applies equally to a three-dimensional bulk system and to a two-dimensional sheet. The distinction enters through Gauss's law. In a bulk system, the Coulomb kernel is local in three dimensions, whereas for a thin sheet embedded in three-dimensional space it becomes nonlocal in the in-plane coordinates. Our main results now follow from an analysis of equations \eqref{eq:phase} and \eqref{eq:charge} for the phase $\phi$ and the density $\nu$ in the presence of the electromagnetic vectorpotential $(A_0, \vecA)$ in different physical situations.

\subsection{Bulk superconductor}

For a three-dimensional bulk superconductor, Gauss's law takes the form
\begin{align}
\nabla^2A_0=\nu.
\end{align}
Combining this relation with Eq.~\eqref{eq:charge}, we obtain the charge-density response
\begin{align}
\nu=\hat{G}_\nu\left(Z_c\rho_0+\alpha n_f\hat\sigma\right)\nabla^2\phi+\hat G_\nu\nabla\hat\sigma\vecxi_L,
\end{align}
where
\begin{align}
\hat G_\nu=\left(\partial_t-n_f\hat\sigma\right)^{-1}
\end{align}
is the charge Green's operator and $\vecxi_L$ the longitudinal projection of $\vecxi_v$.

Substituting both the charge response and the eliminated amplitude fluctuation into the phase equation gives the closed longitudinal equation
\begin{equation}
\begin{split}
&\partial_t\phi-Z_\phi\nabla^2\phi+\Lambda_{KPZ}(\nabla\phi-\vecA)^2\\&+\hat G_\nu\hat\omega_p^2\left(1-\eta\nabla^2\right)\phi+\xi_\phi+\xi_{rest}=0.
\end{split}
\end{equation}
We have introduced
\begin{subequations}
\begin{align}
Z_\phi&=Z_d+\frac{Z_c\lambda_c}{\lambda_d}\\,
\Lambda_{KPZ}&=Z_d\left(\frac{\lambda_c}{\lambda_d}-\frac{Z_c}{Z_d}\right),\\
\eta&=g_d\left(\frac{g_c}{g_d}-\frac{\lambda_c}{\lambda_d}\right),\\
\hat\omega_p^2&=Z_c\rho-\alpha n_f\hat\sigma.
\end{align}
\end{subequations}
Here, $Z_\phi$ is the effective phase-diffusion coefficient, $\Lambda_{KPZ}$ controls the leading nonlinear phase-gradient term, and $\eta$ parametrizes an additional non-equilibrium coupling between phase and density fluctuations. When the effective-equilibrium relations among the reactive and dissipative coefficients eq.~\eqref{eq:thermal_equilibrium} are satisfied, both $\Lambda_{KPZ}$ and $\eta$ vanish.

The term $\xi_{rest}$ collects the colored noise contributions generated by the successive elimination of the amplitude and density fluctuations. Because the original dynamics also contains a Markovian noise contribution, which dominates the long-time correlations, we approximate
\begin{align}
\xi_\phi+\xi_{rest}\simeq\bar\xi_\phi,
\end{align}
where $\bar\xi_\phi$ is again taken to be Gaussian white noise.

Multiplying the phase equation by $\hat G_\nu$
gives
\begin{align}
\partial_t^2\phi&+\left(-n_f\hat\sigma-Z_\phi\nabla^2\right)\partial_t\phi-\left(n_f\hat\sigma Z_\phi+\hat\omega_p^2\eta\right)\nabla^2\phi+\hat\omega_p^2\phi\nonumber\\
&+\left(\partial_t-n_f\hat\sigma\right)\Lambda_{KPZ}(\nabla\phi-\vecA)^2+\left(\partial_t-n_f\hat\sigma\right)\bar\xi_\phi=0.
\end{align}

The Gaussian collective-mode spectrum follows from the linear part of this equation. We consider the strong-damping regime of the normal carriers, in which their relaxation is fast compared with the plasmon dynamics. In this limit,
\begin{align}
\hat\sigma\simeq-\gamma^{-1}.
\end{align}
The plasmon dispersion is then
\begin{align}
\omega(\vecq)^2=Z_c\rho_0^2+\alpha\frac{n_f}{2\gamma_n}-\frac{n_f^2}{4\gamma_n^2}+\bar v^2\vecq^2+Z_\phi^2\vecq^4,
\end{align}
with
\begin{align}
\bar v^2=\frac{Z_\phi n_f}{2\gamma_n}+\eta\left(Z_c\rho_0+\alpha n_f\gamma_n^{-1}\right).
\end{align}
The corresponding decay rate is
\begin{align}
d(\vecq)=\frac{n_f}{\gamma_n}+Z_\phi\vecq^2.
\end{align}
Thus, as in an equilibrium bulk superconductor, the longitudinal collective mode remains gapped at vanishing momentum. The non-equilibrium couplings modify its gap, damping, and dispersive corrections but do not remove the Anderson-Higgs mechanism.

We next consider the transverse electromagnetic response. The transverse component of the normal-carrier velocity is
\begin{align}
\vecv_T=-\hat\sigma\left(\partial_t+\alpha\right)\vecA+\hat\sigma\vecxi_T.
\end{align}
The total transverse current therefore becomes
\begin{align}\label{eq:transversal_current}
\vecj_T=-\left[Z_c\rho_0-n_f\hat\sigma(\alpha+\partial_t)\right]\vecA+n_f\hat\sigma\vecxi_T.
\end{align}
Combining this expression with the transverse Maxwell equation gives
\begin{align}
\left(\partial_t^2-n_f\hat\sigma\partial_t-c^2\nabla^2+Z_c\rho_0-\alpha n_f\hat\sigma\right)\vecA+n_f\hat\sigma\vecxi_T=0.
\end{align}
In the strong-damping limit $\hat\sigma\simeq-\gamma^{-1}$, the transverse photon acquires the usual Meissner gap. At zero momentum, this gap coincides with the plasmon pole, as required by the Anderson-Higgs mechanism. The corresponding inverse penetration depth is
\begin{align}
\xi_L^{-1}=\sqrt{Z_c\rho_0-\alpha\gamma_n^{-1}n_f
}.
\end{align}
The driven-dissipative state therefore exhibits static magnetic-field expulsion despite the finite-frequency rotation of its order parameter.

\subsection{Two-dimensional sheet}

We next consider a two-dimensional superconducting sheet embedded in three-dimensional vacuum. The elimination of the amplitude fluctuation remains unchanged, and Eq.~\eqref{eq:eliminate_amplitude} continues to apply. The difference arises from the electromagnetic field outside the sheet, which must first be solved in the surrounding vacuum.

We take the superconducting sheet to lie in the $xy$ plane at $z=0$. After Fourier transforming with respect to the in-plane coordinates, the scalar potential away from the sheet has the form
\begin{align}
A_0(\vecq,z\neq0)=e^{-|\vecq|z}A_0(\vecq).
\end{align}
Integrating Poisson's equation across the sheet and imposing continuity of $A_0$ gives
\begin{align}
2|\vecq|A_0=\nu.
\end{align}
The Coulomb kernel is therefore nonlocal in the in-plane coordinates.

Combining this constraint with the continuity equation gives
\begin{align}
\nu=G^{2d}_\nu\left(Z_c\rho_0+\alpha n_f\hat\sigma\right)\vecq^2\phi+G_\nu^{2d}\vecq\hat\sigma\vecxi_L,
\end{align}
where
\begin{align}
\hat G_\nu^{2d}=\left(\partial_t-|\vecq|n_f\hat\sigma\right)^{-1}.
\end{align}
Substitution into the phase equation gives
\begin{equation}
\begin{split}
&\partial_t\phi-Z_\phi\vecq^2\phi+\Lambda_{KPZ}(\vecq\phi-\vecA)^2\\
&+\hat G^{2d}_\nu\hat\omega_p^2\left(|\vecq|-\eta\vecq^2\right)\phi+\xi_\phi+\xi_{rest}=0.
\end{split}
\end{equation}

Multiplying by the inverse charge Green's operator, transforming to momentum space, and again taking the strong-damping limit $\hat\sigma\simeq-\gamma^{-1}$ yields
\begin{equation}
\begin{split}
&\partial_t^2\phi+\frac{n_f}{\gamma}|\vecq|\partial_t\vecphi+\hat\omega_p^2|\vecq|\phi\\
&-\frac{n_f\Lambda_{KPZ}}{\gamma}|\vecq|(\vecq\phi-\vecA)^2+\frac{n_f}{\gamma}|\vecq|\bar\xi_\phi=0,
\end{split}
\end{equation}
where terms that are subleading in momentum have been omitted.

The Gaussian plasmon dispersion therefore satisfies
\begin{align}
\omega(\vecq)^2=\left(Z_c\rho_0+\frac{\alpha n_f}{\gamma}\right)|\vecq|+O(\vecq^2),
\end{align}
and the decay rate is
\begin{align}
d(\vecq)=\frac{n_f}{\gamma}|\vecq|.
\end{align}
As in an equilibrium two-dimensional superconductor, the plasmon is gapless and has the characteristic square-root dispersion
\begin{align}
\omega(\vecq)\propto|\vecq|^{1/2}.
\end{align}
At the same time, its damping vanishes linearly in $|\vecq|$. The ratio of the damping rate to the plasmon frequency therefore vanishes as $|\vecq|^{1/2}$ in the long-wavelength limit, and the mode remains underdamped.

The transverse gauge fluctuations can be treated analogously. Solving the electromagnetic field in the surrounding three-dimensional vacuum and integrating across the sheet gives
\begin{align}
\sqrt{\omega^2-c^2\vecq^2}\vecA=\vecj_T.
\end{align}
The transverse current is still given by Eq.~\eqref{eq:transversal_current}. In the static limit, $\omega=0$, one obtains
\begin{align}
\left(c|\vecq|+Z_c\rho_0+\frac{\alpha n_f}{\gamma}\right)\vecA=0.
\end{align}
Thus, a two-dimensional driven-dissipative superconductor exhibits the same form of nonlocal magnetic screening as its equilibrium counterpart. The screening is characterized by a Pearl length,
\begin{align}
l=\frac{c}{Z_c\rho_0+\alpha n_f\gamma^{-1}},
\end{align}
rather than by a bulk penetration depth.

\subsection{Role of interactions and fluctuations}

The principal signature of the non-equilibrium steady state in the effective phase dynamics is the appearance of the KPZ-type nonlinearity proportional to $\Lambda_{KPZ}$. At the Gaussian level, the noise and dissipative coefficients determine the bare response and correlation functions. In an effective-equilibrium theory, these quantities obey fluctuation-dissipation relations. Away from the effective-equilibrium manifold, the nonlinear phase-gradient term and the coupling $\eta$ provide explicit measures of non-equilibrium behavior.

For a driven-dissipative superfluid in two dimensions, the KPZ nonlinearity has a drastic infrared effect. It destabilizes the Gaussian theory and drives the system toward a strongly interacting KPZ fixed point. At this fixed point, the dynamics exhibits anomalous scaling and violates equilibrium fluctuation-dissipation relations at long wavelengths, making the non-equilibrium nature of the condensate directly manifest.

The situation is different for a driven-dissipative superconductor. In three dimensions, the plasmon and transverse photon are both gapped by the Anderson-Higgs mechanism. The gap cuts off potential infrared singularities, so the KPZ interaction does not qualitatively modify the long-wavelength conductivities or susceptibilities. Its effects instead appear in nonlinear processes, such as the scattering of a plasmon into transverse photons.

In two dimensions, the plasmon remains gapless but obeys the nonanalytic dispersion $\omega\propto|\vecq|^{1/2}$. This modified Gaussian dynamics changes the scaling dimension of the nonlinear phase-gradient interaction. Simple power counting shows that the KPZ-type nonlinearity does not generate the infrared singularity familiar from a two-dimensional driven-dissipative superfluid. It therefore does not destabilize the Gaussian fixed point. Consequently, the long-wavelength plasmon spectrum and magnetic response remain controlled by the Gaussian theory, even though their coefficients retain information about the non-equilibrium steady state.

\section{AC Josephson effect without voltage bias}
\label{sec:josephson}

\begin{figure}
    \centering
    \includegraphics[width=0.8\linewidth]{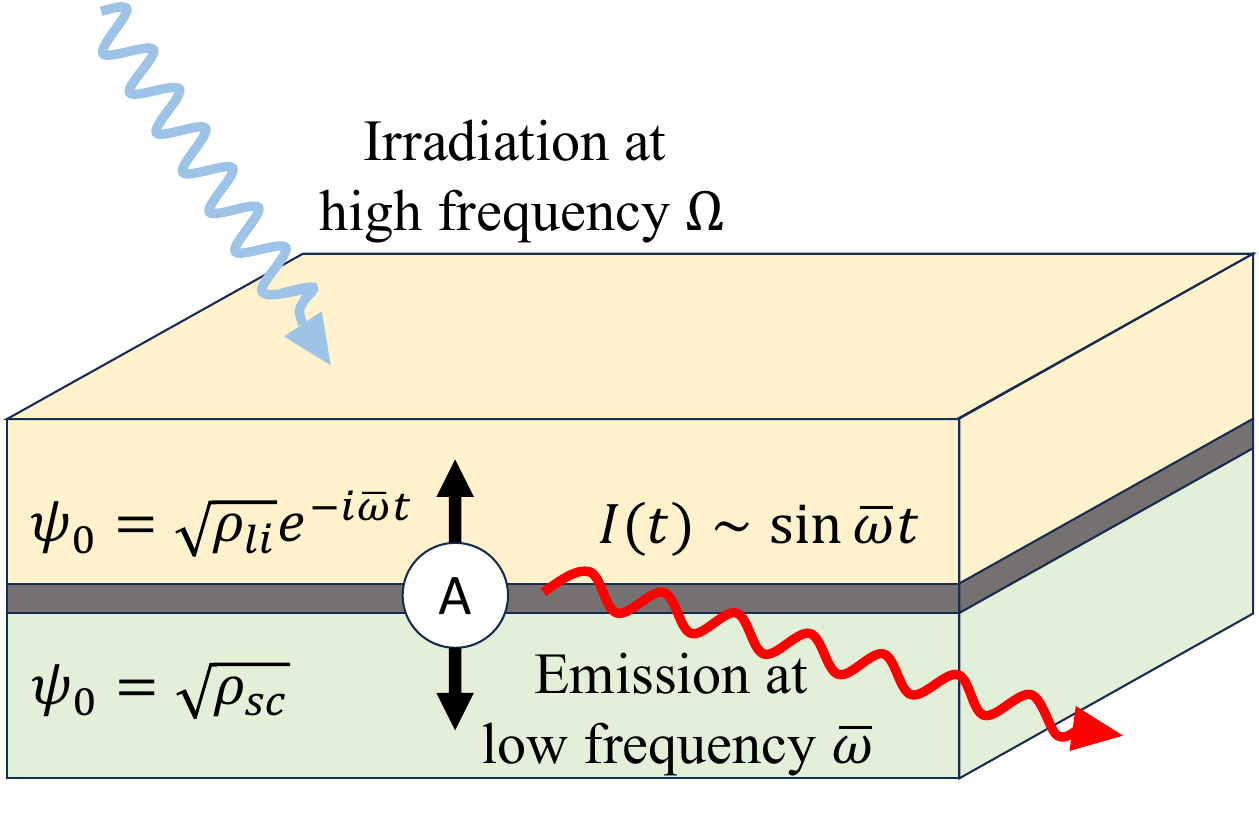}
    \caption{Schematic visualization of photo-induced AC Josephson effect. Upper layer becomes superconducting once driven by a suitable high frequency laser at the bath temperature $T_b$. Lower layer is a high-$T_c$ compound that is naturally superconducting at $T_b$. The resulting phase rotation of the driven-dissipative superconductor results in an AC Josephson current across the junction without any bias applied. This leads to detectable radiation at a frequency $\bar\omega$ lower than and incommensurate with the laser frequency $\Omega$.}
    \label{fig:josephson}
\end{figure}

The prior considerations show that the slow rotation of the order parameter at $\bar\omega$ does not qualitatively alter the plasmon spectrum and Meissner effect with respect to the equilibrium scenario. There is a static magnetic field expulsion as well as a superconducting response despite the order parameter being time dependent. This can be simply rationalized by that fact that the time dependence of the order parameter is fully contained by its phase. The phase of the superconducting order parameter is however not gauge invariant and one can absorb a constant angular velocity completely in a shift of the longitudinal gauge field $A_0$.

The slow rotation is however not unphysical, it becomes observable at a Josephson junction between a photo-induced superconductor and an equilibrium superconductor. The phase difference $\Delta\phi$ between both superconducting compounds is gauge invariant and it oscillates perpetually
\begin{align}
    \Delta\phi=\phi_0+\bar\omega t.
\end{align}
This means that in stark distinction to the equilibrium case there is an AC Josephson current flowing across the junction without applying a bias voltage
\begin{align}
    I(t)=I_c\sin{(\phi_0+\bar\omega t)}.
\end{align}
This leads to radiation at the frequency $\bar\omega$, which is lower and in general incommensurate with the drive frequency $\Omega$. 

The junction should consist of two materials with different critical temperatures $T_1<T_2$. Furthermore, material 1 should display photoinduced superconductivity at a bath temperature $T_1<T_{bath}<T_2$. For instance $K_3C_{60}$ has a $T_c$ of $20K$ but shows light-induced superconductivity up to temperatures of $100K$ \cite{mitrano2016}. Then, such a junction would show no Josephson effect without the drive. As soon as the material 1 (e.g. $K_3C_{60}$) is pumped sufficiently and it becomes a driven-dissipative superconductor, there is an oscillating current at the junction leading to radiation. Observing this radiation would confirm theories of a finite-frequency superconductor. In a superconductor oscillating in resonance with the drive this radiation would be at integer multiple of the drive frequency, as proposed in \cite{Dai22}, while a driven-dissipative SC would radiate at a lower, incommensurate frequency. Furthermore this frequency is pump power dependent. This is schematically visualized in Fig.~\ref{fig:josephson}

\section{Discussion and outlook}

We have developed a phenomenological description of photo-induced superconductivity based on a dissipative instability of the superconducting order parameter. Our central result is that, despite its intrinsic slow phase rotation, the resulting non-equilibrium state behaves as an electromagnetic superconductor: it expels static magnetic fields and exhibits the collective electromagnetic response associated with superconducting order. At the same time, the phase rotation provides a clear experimental signature that distinguishes the driven-dissipative state from both an equilibrium superconductor and a pairing field locked to the external drive. A Josephson junction between a driven-dissipative superconductor and an equilibrium superconductor supports an AC current at zero applied voltage and therefore emits radiation at the intrinsic rotation frequency of the condensate. This frequency is expected to be pump-power dependent and, in general, lower than and incommensurate with the optical drive.

The detailed electromagnetic response depends on dimensionality. In a three-dimensional bulk system, the Anderson--Higgs mechanism produces a gapped plasmon and a finite London penetration depth. In a two-dimensional superconducting sheet embedded in three-dimensional space, the nonlocal Coulomb interaction instead gives rise to the characteristic square-root plasmon dispersion and magnetic screening governed by a Pearl length. These results show that the finite-frequency dynamics of the order parameter does not qualitatively alter the conventional long-wavelength signatures of superconductivity. The precise collective-mode frequencies, damping rates, and screening scales depend on the phenomenological coefficients, but the coexistence of conventional electromagnetic superconducting response and zero-bias Josephson radiation is a robust feature of the driven-dissipative scenario.

The continuum theory also provides a starting point for studying the critical behavior of photo-induced superconductivity. Dissipative ordering transitions can be governed by genuinely non-equilibrium universality classes that are distinct from their equilibrium counterparts \cite{daviet2024}. An important open question is how this critical behavior is modified by the simultaneous presence of a conserved charge density and dynamical gauge fields. In particular, it remains to be determined whether the transition is controlled by a new non-equilibrium fixed point or whether gauge fluctuations suppress the fluctuation-dominated regime.

A complementary direction is to derive the dissipative instability from microscopic models of optically driven materials. Such an analysis should identify the pumped reservoir, the relaxation processes that transfer spectral weight into the pairing sector, and the conditions under which the resulting gain overcomes the intrinsic damping. It would also allow the phenomenological coefficients of the continuum theory to be related to pump frequency, pump intensity, temperature, and material parameters. This step is essential for assessing which of the experimentally studied compounds are plausible candidates for driven-dissipative superconductivity.

Finally, the present framework can be extended to unconventional superconducting order parameters. For example, in a driven $d$-wave superconductor, nodal quasiparticles may modify the effective damping, the electromagnetic response, and the stability of the rotating state. Multicomponent order parameters may also support more complex non-equilibrium trajectories than a uniform phase rotation. 

Establishing this phenomenology experimentally would show whether photo-induced superconductivity can be understood as an ordered state stabilized not by an effective free-energy minimum, but by a genuinely non-equilibrium balance of gain and loss.
  
\acknowledgements

We thank Andrea Cavalleri, Eugene Demler, Patrick Lee, Marios Michael,  Matteo Mitrano and Pavel Nosov for valuable discussions. This research was supported by NSF Grant DMR-2245246. The Flatiron Institute is a division of the Simons Foundation. CZ was supported by by the Deutsche Forschungsgemeinschaft (DFG, German Research Foundation) project number 570906600.

\bibliography{dissipativeSCbiblio}

\end{document}